# Targeted Sub-attomole Cancer Biomarker Detection based on Phase Singularity 2D Nanomaterial-enhanced Plasmonic Biosensor

Yuye Wang[1,2], Shuwen Zeng[2,3*], Aurelian Crunteanu[2], Zhenming Xie[1], Georges Humbert[2], Libo Ma[4], Yuanyuan Wei[1], Aude Brunel[5], Barbara Bessette[5], Jean-Christophe Orlianges[2], Fabrice Lalloué[5], Oliver G Schmidt[4], Nanfang Yu[3], Ho-Pui Ho[1*]

[1]Department of Biomedical Engineering, The Chinese University of Hong Kong, Shatin, New Territories, Hong Kong.

[2]XLIM Research Institute, UMR 7252 CNRS/University of Limoges, 123, Avenue Albert Thomas, Limoges, France.

[3]Department of Applied Physics and Applied Mathematics, Columbia University, New York, USA.

[4]Institute for Integrative Nanosciences, IFW Dresden, Helmholtzstr. 20, Dresden, Germany

[5]EA3842- CAPTuR, GEIST, Faculty of medicine, University of Limoges, 2 rue du Dr Marcland, Limoges, France

*Corresponding author. E-mail: aaron.ho@cuhk.edu.hk, zeng@xlim.fr

**Highlights**

- In this work, a zero-reflection induced phase singularity is achieved through precisely controlling the resonance characteristics using 2D nanomaterials.

- An atomically thin nano-layer having a high absorption coefficient is exploited to enhance the zero-reflection dip, which has led to the subsequent phase singularity and thus a giant lateral position shift.

- We have improved the detection limit of low molecular weight molecules by more than 3 orders of magnitude compared to current state-of-art nanomaterial-enhanced plasmonic sensors.

**TOC**

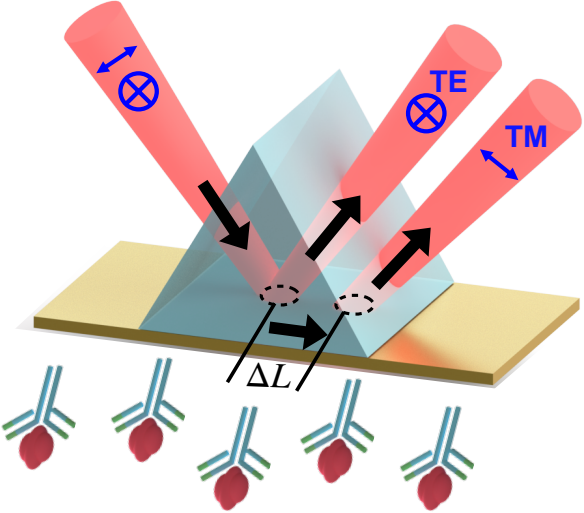

**Abstract**

Detection of small cancer biomarkers with low molecular weight and a low concentration range has always been challenging yet urgent in many clinical applications such as diagnosing early-stage cancer, monitoring treatment and detecting relapse. Here, a highly enhanced plasmonic biosensor that can overcome this challenge is developed using atomically thin two-dimensional (2D) phase change nanomaterial. By precisely engineering the configuration with atomically thin materials, the phase singularity has been successfully achieved with a significantly enhanced lateral position shift effect. Based on our knowledge, it is the first experimental demonstration of a lateral position signal change > 340 μm at a sensing interface from all optical techniques. With this enhanced plasmonic effect, the detection limit has been experimentally demonstrated to be $10^{-15}$ mol·L$^{-1}$ for TNF-α cancer marker, which has been found in various human diseases including inflammatory diseases and different kinds of cancer. The as-reported novel integration of atomically thin $Ge_2Sb_2Te_5$ (GST) with plasmonic substrate, which results in a phase singularity and thus a giant lateral position shift, enables the detection of cancer markers with low molecular weight at femtomolar level. These results will definitely hold promising potential in biomedical application and clinical diagnostics.

# 1 Introduction

To promote early-stage diagnostics of a variety of human diseases, the detection of specific biomarkers at extremely low concentration levels has attracted a lot of attention over these years [1-3]. Among the various biomarkers, tumor necrosis factor TNF-α has been studied intensively and proved to play a central role in mammalian immunity and cellular homeostasis [4, 5]. TNF-α is a key mediator in pro-inflammatory responses. It is also involved in various cell activities including cellular communication, cell differentiation and cell death [6]. These regulatory functions have made TNF-α an important biomarker for monitoring a variety of human diseases, including inflammatory disorder such as bowel disease [7], osteoarthritis [8, 9], rheumatoid arthritis [10] as well as malignant tumors such as oral [11] and breast cancer [12, 13], etc. The detection of biomarkers, particularly TNF-α, is certainly of great significance to healthcare efficacy through early diagnosis and monitoring of life-threatening diseases. However, the concentration level of this biomarker is extremely low, typically ~20 pg/mL in a healthy human [14, 15], thus making detection of this molecule a significant challenge. Moreover, the molecular weight of TNF-α (~17 kDa) is one order of magnitude lower than many common biomarkers such as argonaute proteins [16] (~100 kDa) and carcinoembryonic antigen [17] (~180 kDa), which further increases the level of difficulty.

Conventional detection techniques include gel electrophoresis, enzyme-linked immunosorbent assay (ELISA) and fluorescence-based detection [18, 19], etc. However, traditional detection methods are usually time-consuming and require complex operations with adequate transducing elements such as fluorescent dyes or expensive enzymes [20]. Over these years, some cutting-edge technologies including electrochemical approaches [21] and microfluidic-based approaches [22], have been developed to enable more effective and sensitive cancer marker detection. More recently, several research groups have exploited mass spectrometry (MS) based on optimized nanoparticles as matrix materials to enable fast detection with high selectivity and sensitivity [23-25]. This MS-based detection can directly detect small metabolites in human serum without any purification in advance [26, 27]. However, bulk instrumentation is needed in this case, which largely limits the application in on-site detection. As a result, there is still much room for improvement in terms of low-cost, convenient and ultra-sensitive biodetection as well as point-of-care diagnostics.

Due to its label-free, real-time and low-cost detection merits, the surface plasmon sensing technique has been exploited for a range of biosensing applications [28-30]. However, for target analytes that have low molecular weights (less than 400 Dalton), plasmonic sensors still face the challenges to compete in terms of detection sensitivity [31]. Compared to traditional plasmonic biosensor designs based on angle [32, 33] or wavelength [34-36] interrogation, the phase detecting approach has been shown to improve the sensitivity limit by a few orders of magnitude [37, 38]. Unlike the moderate change in intensity or wavelength, the optical phase experiences an abrupt change when the reflection reaches almost zero. Furthermore, the phase detection can also provide lower noise and allows for versatile signal-processing possibilities [39]. Recently, with increasing efforts to explore the concept of "point of darkness", which represents the zero-reflection point, several research groups have exploited plasmonic metamaterials that exhibit topologically protected darkness for the design of biosensing devices offering radically enhanced sensitivity due to extremely steep phase variations [37, 40, 41]. Further, a higher order of the phase signal, the lateral position shift, may bring further improvement in sensitivity. Since zero reflection can lead to a singular behavior of the phase in Fourier space, the resulted sharp phase jump will then induce a giant lateral position shift, making it an excellent choice for sensing biomolecules at very low concentrations [42, 43].

In this work, we have significantly enhanced the performance of a SPR biosensing platform by adding an atomically thin phase change material to induce a giant lateral position shift called Goos–Hänchen (GH) shift, which in turn leads to the detection of TNF-α cancer biomarkers at sub-attomole level. We tuned the atomically thin $Ge_2Sb_2Te_5$ (GST) nanomaterials, which have a high absorption rate [44] in visible and near-infrared wavelengths, to achieve zero-reflection at plasmon resonance. The zero-reflection phenomenon can result in a strong phase singularity. This zero-reflection induced phase singularity is known to be challenging to achieve in previous plasmonic nanostructures [37, 45]. The higher order mode of the phase signal, i.e., the lateral position shift, was found to be much larger than other signal modalities reported in recent years [46-49]. The plasmonic sensing device reported herein exhibits a detection limit of $10^{-15}$ mol/L (1 fM) for TNF-α biomarkers and $10^{-14}$ mol/L (10 fM) for small biotin molecules (MW=244.31 Da). This sub-attomole detection level is a significant improvement compared to other SPR designs [50, 51]. The maximum experimental lateral position shift triggered in our device is 341.90 μm, which to our best knowledge is the largest value ever reported. For detecting small biomolecules and cancer markers at femto-molar concentration levels, the sensing signal was around 10 μm, which is quite readily measured by an interrogation setup. In summary, the proposed scheme has been shown to be capable of sensing extremely small refractive index (RI) changes, which is of great interest for label-free biosensing sensing application and clinical diagnostics.

## 2 Experimental methods

### 2.1 Methodology of the enhanced plasmonic sensor

We have significantly enhanced the sensitivity of a plasmonic sensing system by engineering an atomically thin GST layer to the plasmonic substrate. The designed atomically thin GST-on-Au plasmonic biosensing scheme was based on the Kretschmann configuration as shown in **Figure 1** (a). For the sensing layer design, an atomically thin layer of 2D GST material is uniformly deposited on the top of the Au thin film. The GST phase change materials are known to have a higher absorption rate in visible and near-infrared regions than some other 2D materials such as graphene, $MoS_2$ and $WS_2$ [52-54]. As a result, with careful optimization of the thickness of GST layer, a condition very close to zero-reflection can be achieved. Under this near zero-reflection phenomenon, we can observe not only a fast drop in reflected light intensity, but also an extremely sharp phase change at the resonance angle, which can be exploited to greatly enhance the sensitivity based on plasmon resonances. The lateral position shift is a higher order optical signal of the phase singularity (See Equation (3)) and is readily detectable by our plasmonic sensing setup. It is worth noting that only *p*-polarized light will have this giant phase singularity-induced lateral position shift under the plasmonic excitation while *s*-polarized light remains unaffected (Figure 1 (b)). Therefore, *s*-polarized light serves as a reference here to eliminate environmental disturbances and can be used to significantly improve the signal to noise ratio of the measurement. The differential signal acquired by the position sensor between the *p* and *s* polarizations can provide high signal-to-noise measurements. The largest lateral position shift is achieved when the sharpest phase change occurs, which corresponds to the minimum reflectance point within the surface plasmon resonance dip. Our theoretical results have shown that the lateral position shift is inversely correlated with the reflectivity (See Supporting information Figure S1). In order to achieve the largest lateral position shift, we have enhanced the absorption characteristics to near zero-reflectance ($\sim 10^{-6}$) by adding an atomically thin 2D phase change material. Based on the theoretical model, the optimized thicknesses of the 2 nm GST layer can lead to minimum reflectance, i.e., maximum sensitivity. The lateral position shift of both Au-only substrate and our atomically thin GST-on-Au substrate was simulated. It is clearly shown in Figure 1 (c) that the addition of atomically thin GST material can lead to an extremely giant maximum lateral position shift up to 2107.33 μm at the resonance angle, which is nearly 100 times larger than that associated with the case of using Au-only substrate. Here, we also demonstrated the extreme singularities in phase of the reflected light enhanced by GST-on-Au substrate in Figure 1 (d). Compared to Au-only substrate, the phase change tends to be much sharper, which indicates the radically enhanced sensitivity based on our GST-on-Au substrate. Moreover, the electric field distribution on this 2D GST-on-Au sensing substrate at the resonance angle was also studied using finite element analysis (FEA) (COMSOL Multiphysics 5.2) (Figure S2 (a), Supporting Information). The large electric field enhancement at the sensing interface also demonstrated the enhanced plasmonic resonance. These results have shown that the atomically thin 2D GST layer will offer a superior sensitivity enhancement.

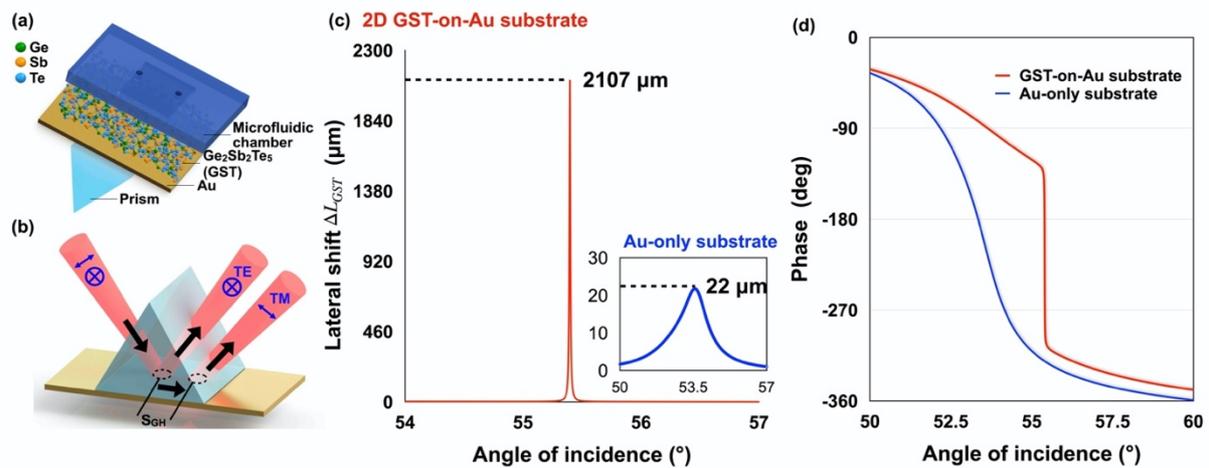

**Fig. 1** (a) Schematic of the sensing substrate based on GST-gold metastructures. (b) Schematic diagram of the giant lateral position shift induced by the GST-on-Au substrate. Simulation results of (c) lateral position shift and (d) optical phase signal change based on 2D GST-on-Au substrate and Au-only substrate.

## 2.2 Optimization of the 2D GST thickness on the plasmonic sensing substrate

In a Kretschmann configuration, the reflection coefficient of two adjacent layers takes the form of

$$r_{i,i+1} = \frac{Z_i - Z_{i+1}}{Z_i + Z_{i+1}} \quad (1)$$

where $Z_{ip} = \frac{\varepsilon_i}{k_i}$ for $p$ polarization and $Z_{is} = k_i$ for $s$ polarization. $\varepsilon_i$ represents the complex dielectric constants of the i-th layer and $k_i = k_0\sqrt{\varepsilon_i - \varepsilon_1 sin^2\theta_c}$, in which $k_0$ is the wave vector of the optical wave in free space and $\theta_c$ is the incident angle.

For a sensing substrate with m layers, we have

$$r_{m-2,m} = \frac{r_{m-2,m-1} + r_{m-1,m}\exp(2ik_{m-1}d_{m-1})}{1 + r_{m-2,m-1}r_{m-1,m}\exp(2ik_{m-1}d_{m-1})} \quad (2)$$

Then we subsequently calculated $r_{m-3,m}$, $r_{m-4,m}$ ... until we get $r_{1,m}$, which is the reflectivity of the substrate based on this metastructure.

From Fresnel's equations, the complex reflection coefficients can be expressed as $r_{p(s)} = |r_{p(s)}|\exp(i\phi_{p(s)})$ for $p$ and $s$ polarizations in which $\phi_{p(s)}$ represents the phase of both polarizations. According to the stationary phase approach [55], the lateral position shift the lateral position shift represents the higher order mode of the phase signal [41, 56, 57] and can be determined through the following equation [58]:

$$\Delta L = -\frac{1}{k_0}\frac{\partial \phi}{\partial \theta} \quad (3)$$

If we determine the reflection coefficients through the standard characteristic matrix approach [59, 60], the lateral position shift can also be expressed as:

$$\Delta L = -\frac{\lambda}{2\pi|r_{p(s)}|^2}\left(Re(r_{p(s)})\frac{dIm(r_{p(s)})}{d\theta} - Im(r_{p(s)})\frac{dRe(r_{p(s)})}{d\theta}\right) \quad (4)$$

where *Re* is the real part and *Im* is the imaginary part. Based on the above equations, the optimized thickness of 2D GST layer for achieving the minimum reflectance and maximum lateral position shift can therefore be calculated.

**2.3 Materials and methods**

**Chemicals.** Glycerol, absolute ethanol, lyophilized Bovine Serum Albumin (BSA) powder, lyophilized biotin powder, and (3-Aminopropyl) trimethoxysilane (APTMS), Tumor necrosis factor-alpha (TNF-α) antigen powder, Monoclonal Anti-TNF antibody were purchased from Sigma Aldrich, France. Colloidal solutions with uniformly dispersed single $MoS_2$ nanosheets were purchased from Ossila Ltd, UK.

**Device fabrication.** $Ge_2Sb_2Te_5$ (GST) and Au layers of the sensing substrate were fabricated by MP300 DC magnetron sputtering equipment (Plassys-Bestek, France) using stoechiometric 2" diameter GST and Au (high purity targets 99.99%, Neyco Vacuum & Materials, France) targets on glass substrates respectively. The deposition chamber was pumped down to $2 \times 10^{-6}$ mb prior to the deposition. The deposition took place under Ar atmosphere (60 sccm flow rate) at $5 \times 10^{-3}$ mb and $1 \times 10^{-2}$ mb partial pressures respectively, using DC magnetron powers between 25 and 55 W.

**Biosample preparation.** Glycerol solutions with concentrations from 1–5% (weight ratio) were prepared. All the solvents used here are deionized water. BSA and biotin solutions with concentrations from 10 fM to 10 μM were prepared by serial dilution (1:100). Antibody solutions with 10 pM concentration were diluted several times from originally 0.5 mg/ml Monoclonal Anti-TNF solutions. TNF-α human lyophilised powder was reconstituted to a concentration of 0.5 mg/ml and further diluted to cancer marker solutions with concentrations from 1 fM to 1 nM.

**Surface functionalization.** The 2D GST-on-Au substrate was first dipped in ethanol and deionized water, followed by drying under nitrogen for clean usage. The GST-on-Au substrate was first immersed in 1mmol/L linker for 30 min to ensure efficient binding of biomolecules as well as cancer markers onto the substrate. Biomolecules at various concentration levels were then injected into the microfluidic chamber and incubated for around 20 min at room temperature for lateral position shift signal collection.

## 3 Results and discussion

### 3.1 Evaluation of the sensor device

A schematic illustration of our experimental setup is shown in **Figure 2** (a). The incident light beam from a He-Ne laser is split into *p*-polarized and *s*-polarized light beams through a polarized beam splitter. An optical chopper is used to ensure that only one of the polarizations can reach the sensing substrate in a fixed period time. A high refractive index prism was mounted on a translation stage with the beam fixed at the surface plasmon resonance dip angle for achieving maximum lateral position shift. The sensing substrate is integrated with a microfluidic chamber to realize convenient transportation of sample solutions using a syringe pump. The real-time positions of the *p*-polarized and *s*-polarized reflected light beams are recorded by a lateral position sensing detector. The lateral position signals are then collected through a data acquisition card and analyzed using a LABVIEW plus MATLAB program. During the biosensing processes, when sample liquids are pumped into the microfluidic chamber, we are able to conduct real-time measurement of the phase singularity-related lateral position shift, where the lateral shifts are induced by the binding of target molecules to the sensing surface.

To evaluate the performance of the 2D GST-on-Au sensing substrate, the angular scanning reflectivity spectra of 2D GST-on-Au and Au-only substrate in air was measured. The experimental results show good agreement with theoretical calculations, which confirms the reliability of our device and serves as a good calibration for the assessment of sensing performance. As shown in Figure S4, the presence of GST material clearly leads to a deeper resonance dip (minimum intensity lowered by 50%). The stronger zero-reflection effects also result in a much larger lateral position shift.

Further, as a standard sensor evaluation procedure, we sequentially injected glycerol solutions of different concentration levels into the microfluidic chamber while the reflectivity is monitored. Figure 2 (b) shows the signals acquired under different glycerol concentrations using atomically thin GST-on-Au substrate. The summarized lateral position shifts of glycerol solutions with concentration levels from 1% to 5% for both Au-only substrate and 2D GST-on-Au substrate were also plotted. Experimental results confirm the significant enhancement in lateral position shift associated with the incorporation of atomically thin GST, which is in good agreement with our theoretical calculation. Also, the measured lateral position shift has a linear relationship with glycerol/water weight ratios. The measured signal change for 1% glycerol (0.0012 Refractive Index Unit, RIU) using atomically thin GST-on-Au substrate is 71.38 μm, which corresponds to a sensitivity figure-of-merit of $5.95\times10^4$ μm/RIU. The position of *p*-polarized reflected light changes drastically while *s*-polarized reflected light remains in the same position upon injecting the glycerol/water solutions into the chamber as shown in Figure S5 (Supporting information). The signal is very stable due to the use of differential measurement scheme.

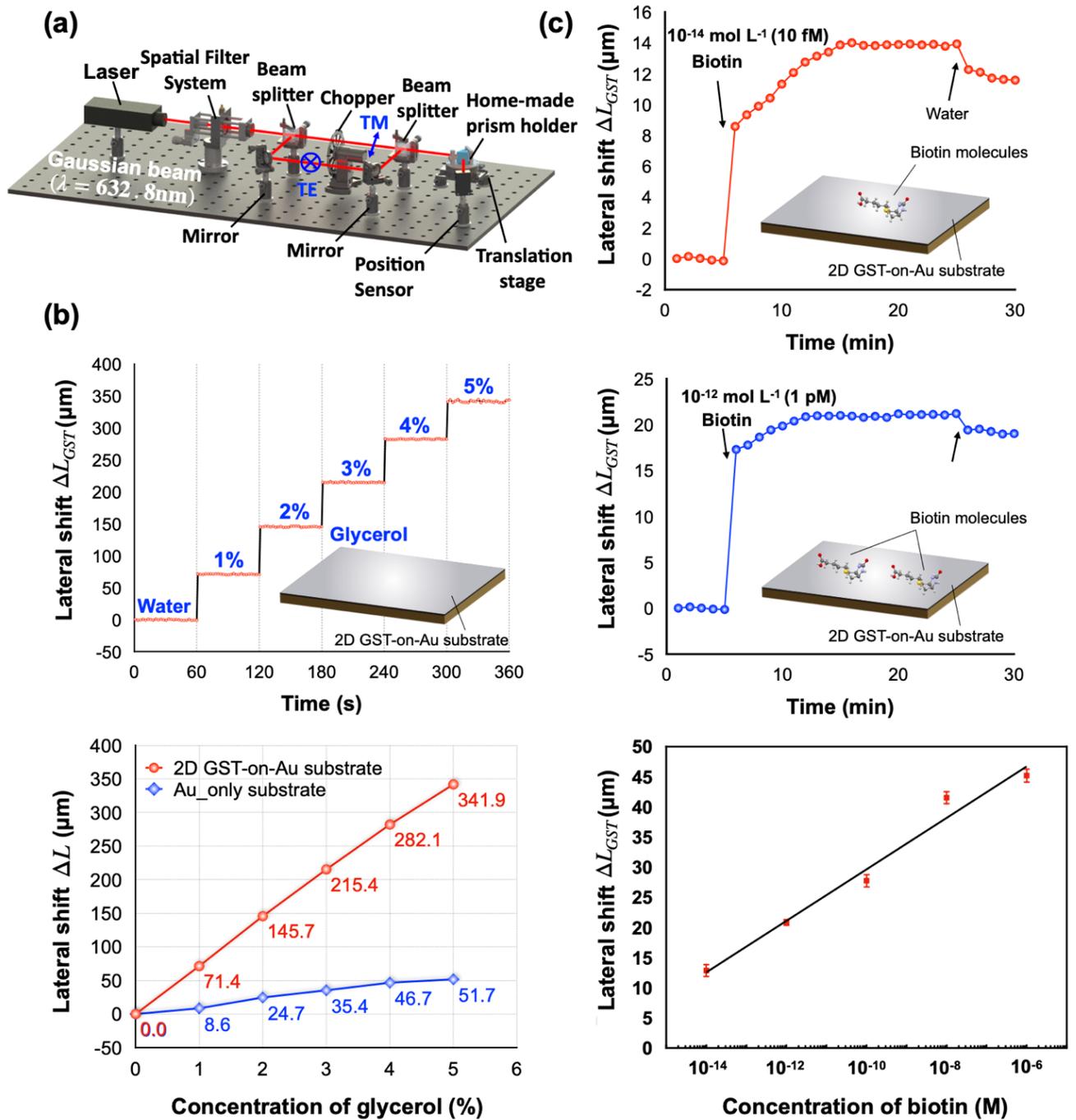

**Fig. 2** (a) Optical setup for measuring differential lateral position shift between *p*- and *s*-polarized light. (b) Evaluation results of measuring glycerol solutions. (c) Real-time detection of biotin molecules based on lateral position shift measurement.

## 3.2 Biomolecule sensing

It is known that the main challenge of optical detection lies in the detection of biomolecules with very low molecular weight (less than 400 Da) such as biotin with a molecular weight of 244 Da. Conventional plasmonic biosensors can only detect biotin molecules with concentration higher than 100 μM [61]. With the recent development of nanomaterial enhanced plasmonic sensors, the detection limit has been lowered to nanomolar [62] or picomolar [63] level. In this work, we are able to further improve the detection limit by more than 3 orders of magnitude. Solutions of the biomolecules with different concentrations ranging from 10 fM to 10 μM were prepared. To ensure efficient binding to the 2D GST-on-Au substrate, we first injected a chemical linker (3-Aminopropyl) trimethoxysilane (APTMS) into the microfluidic chamber to functionalize the sensing substrate with alkoxysilane molecules. This was followed by sequential injection of biotin biomolecules at different molarities to the chamber, which then triggered the corresponding lateral position shifts. Figure 2 (c) shows the detection of biotin molecules based on the measurement of lateral position shifts. The lateral position signal due to biotin binding changed with low speed after a step increase in concentration and saturated at around 14 μm for a biotin concentration of 10 fM. For biotin with a concentration of 1 pM, the lateral shift change shows a similar trend and stabilized at around 20 μm after the saturation time (20 min). This is quite impressive considering the low molecular weight and extremely low concentration level. After each detection run, we flushed away the excess biotin molecules not bound to the substrate. The lateral position shift barely changed, suggesting that most of the targeted molecules have been bound to the substrate. The observed change in lateral position shift is primarily attributed to the binding between the biomolecules and the substrate. The differential lateral position shift signals acquired under biotin concentrations ranging from $10^{-14}$ to $10^{-6}$ M were also summarized, which shows that the lateral shift is linearly proportional to log scale of biotin concentration. Our experimental results have clearly shown that the proposed 2D GST-on-Au biosensor scheme exhibits an ultra-high sensitivity, which is of significance for real-time label-free biosensing.

Besides biotin molecules, we also detected bovine serum albumin (BSA) molecules, which have relatively high molecular weight (66463 Da), at different concentration levels based on the giant lateral position shift. In BSA sensing experiments, the lateral position shift shows an abrupt change immediately after the injection of BSA solutions.

The minimum detectable concentration is estimated to be lower than 10 fM. The lateral position shift has increased to 48.34 μm after a binding event with 10 fM BSA. Solutions of the biomolecules with different concentrations ranging from 10 fM to 10 μM were detected and recorded in Figure S6 (Supporting information) based on lateral position shift measurement, which shows a linear increase in lateral position shifts with increasing BSA concentrations. Signal saturation will start when the concentration goes above $10^{-6}$ mol/L. Given the large lateral position shift at 10 fM levels, for both BSA and biotin, we can assert that this sensor has the capacity of sub-attomole detection sensitivity.

**3.3 Sub-attomole cancer biomarker detection**

Further, to demonstrate the capability of this biosensor for clinical applications, we carried out TNF-α (tumor necrosis factor α) antigen detection using a sandwich immunoassay strategy. In our experiment, we first immersed the sensing surface with capture antibodies. To block non-specific binding in the sensing surface, we flowed BSA molecules, which is widely used as blocking agent [64-66], into the microfluidic chamber in advance to block the unbound sites in the sensing substrate. Then an antigen-containing solution was injected into the microfluidic chamber. A rapidly increasing lateral position shift indicates the specific binding between antigen and antibody molecules. As shown in **Figure 3** (a), our experimental results confirm that the projected detection limit of our atomically thin GST-on-Au plasmonic biosensor can reach $10^{-15}$ M (1 fM) for TNF-α antigen detection. The antigen at a quantity as low as 0.05 attomoles, corresponding to 1 fM in a 50 μl solution, can be measured. To better show the superior sensing capability of the proposed sensing device, a comparison experiment was conducted using Au substrate enhanced by 2D $MoS_2$ nanosheets (Figure 3 (b)).

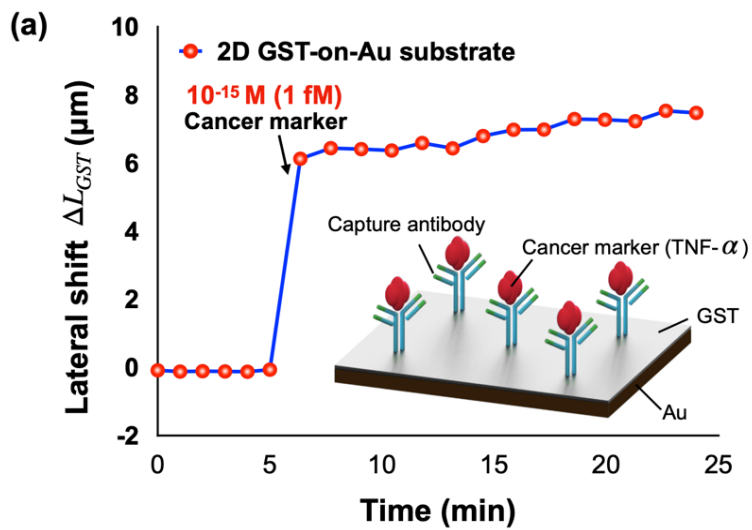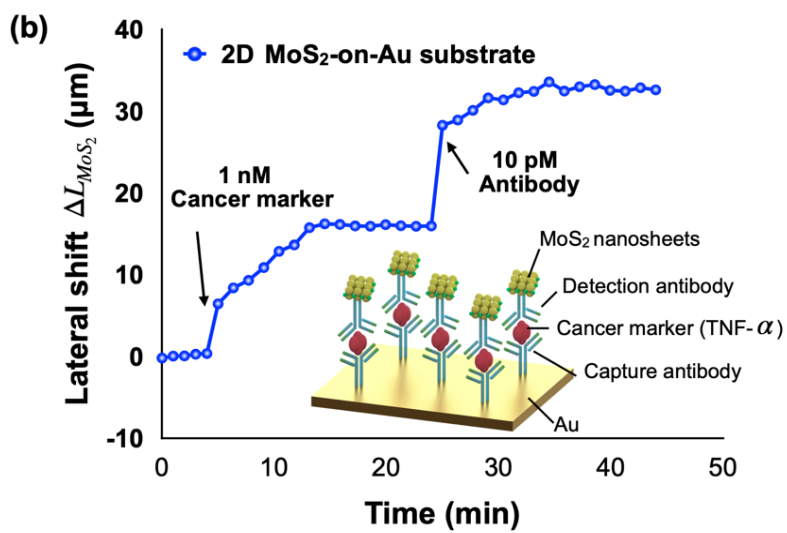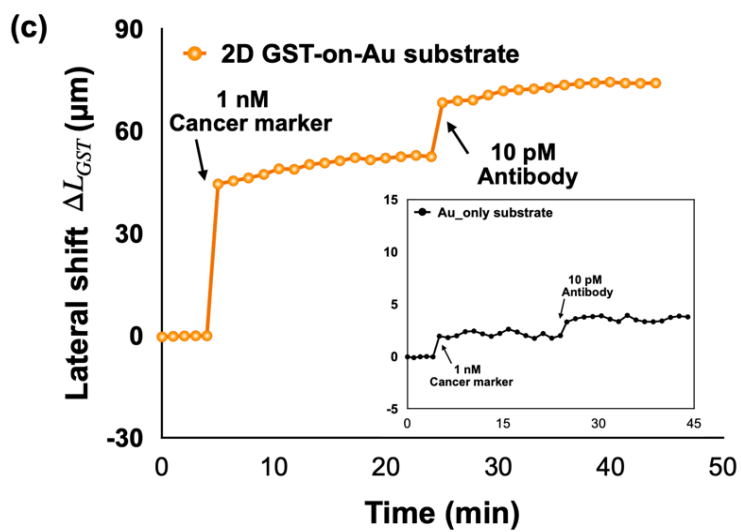

**Fig. 3** (a) Detection of extremely low (1 fM) cancer marker concentration using 2D GST-on-Au substrate. (b) Detection of 1nM cancer marker using 2D MoS$_2$-on-Au substrate. (c) Detection of 1nM cancer marker using 2D GST-on-Au substrate with insert figure showing the performance of Au-only substrate.

In this experiment, we flowed 2D MoS$_2$ nanosheets, which is known as standard 2D materials that are used as the signal amplification nanotags [67-69], into the microfluidic chamber. With the use of 2D MoS$_2$ nanosheets, the lateral position shift signal can be increased to 16.04 μm for 1 nM cancer marker detection. Then with the injection of antibody-containing solution, which binds the antigen at a different epitope than the capture antibody, the lateral position shift will increase again, thus further enhancing the sensing signal. However, even under careful optimization of the 2D MoS$_2$ nanosheets-on-Au substrate, which is much more costly than that case of atomically thin GST, the sensitivity enhancement is still only approximately one-third of our proposed GST-coated substrate. As shown in Figure 3 (c), the lateral shift enhanced by GST-on-Au substrate can reach more than 50 μm after the binding of 1 nM cancer marker biosamples. Under the same configuration, we also conducted experiments on Au-only sensing substrate. Experimental results showed that there is hardly any identifiable lateral position shift change upon injection of the biosamples (insert figure of Figure 3 (c)). In addition to the remarkable sensitivity, the specificity of our sensing device has also been verified through comparing this specific binding process with the non-specific binding between TNF-alpha and BSA molecules. The lateral position shift signal when flowing BSA with a large concentration ($10^5$ times higher than anti-TNF antibody) only increased to 4.37 μm, which is much smaller compared to the signal acquired after flowing monoclonal anti-TNF antibody (16.70 μm), as shown in Figure S7 (Supporting information). The experimental results indicate that our sensing device also has great potential in realizing high-specificity detection in addition to the ultra-high sensitivity.

## 4 Conclusions

In this paper, we present a new plasmonic biosensor with significantly enhanced sensitivity due to the giant lateral position associated with a phase singularity. The sensing substrate is constructed by depositing an atomically thin Ge$_2$Sb$_2$Te$_5$ phase change nanomaterial on gold thin film. The high absorbance of GST material significantly suppresses the reflectivity to a point close to the topological darkness, which yields to extremely steep phase change and further results in a giant lateral position shift with respect to refractive index variations. Refractive index sensing experiments involving the use of glycerol solutions at different concentration levels shows a sensitivity figure-of-merit of $5.95 \times 10^4$ μm/RIU. The maximum detectable lateral position shift is 341.90 μm, which is the highest ever reported in the literature to our best knowledge. Furthermore, we have demonstrated the capability of our device for biosensing using biotin (small molecules) and BSA (large molecules). The experimental detection limit for light molecules (biotin, 244.31 Da) is in the order of $10^{-14}$ M. This is a significant improvement given that most existing SPR system are only achieving pM detection sensitivity limit for low-molecular-weight biomolecules. Our experiments also reveal that atomically GST layer exhibits 3 times better improvement than 2D $MoS_2$ nanosheets, which is commonly used as the signal amplification nanotags. The proposed plasmonic device also shows excellent performance for cancer marker detection. The detection limit has been experimentally demonstrated to be $10^{-15}$ M for TNF-α cancer marker, which is orders of magnitude higher than most label-free detection methods. To summarize, the reported integration of atomically GST layer with plasmonic substrate has been shown to be useful for ultrasensitive biosensing applications with sub-attomole detection limit. It is our view that this label-free, real-time, highly sensitive biosensor has great potential for monitoring chemical and biological reactions with ultra-high sensitivity especially for the clinical diagnostic applications.

## Acknowledgments

We thank Shiyue Liu from School of Life Sciences in The Chinese University of Hong Kong for helpful discussions. This work is supported under the PROCORE - France/Hong Kong Joint Research Scheme (F-CUHK402/19) and by the Research Grants Council, Hong Kong Special Administration Region (AoE/P-02/12, 14210517, 14207419, N_CUHK407/16). This project has received funding from the European Union's Horizon 2020 research and innovation programme under the Marie Sklodowska-Curie grant agreement No. 798916. Y. Wang is supported under the Hong Kong PhD Fellowship Scheme.

Supporting Information

**Targeted Sub-attomole Cancer Biomarker Detection based on Phase Singularity 2D Nanomaterial-enhanced Plasmonic Biosensor**

*Yuye Wang, Shuwen Zeng\*, Aurelian Crunteanu, Zhenming Xie, Georges Humbert, Libo Ma, Yuanyuan Wei, Barbara Bessette, Jean-Christophe Orlianges, Fabrice Lalloué, Oliver G Schmidt, Nanfang Yu, Ho-Pui Ho\**

# Contents

# 1. Theoretical modeling

## 1.1. Relationship between lateral position shift and reflectivity

In our simulation analysis, we calculated the reflection coefficient and lateral position shift under different incident angle using both Au-only substrate and atomically thin GST-on-Au substrate. To demonstrate the relationship between the reflectivity and lateral position shift more clearly, a plot showing their negative correlation was drawn through Fresnel equations and transfer matrix method (TMM) and calculated with a MATLAB programming (**Figure S1**). This explains more explicitly that the maximum lateral position shift is achieved at the minimum reflectivity. Therefore, it is very essential to enhance the zero-reflection effects of the sensing substrate.

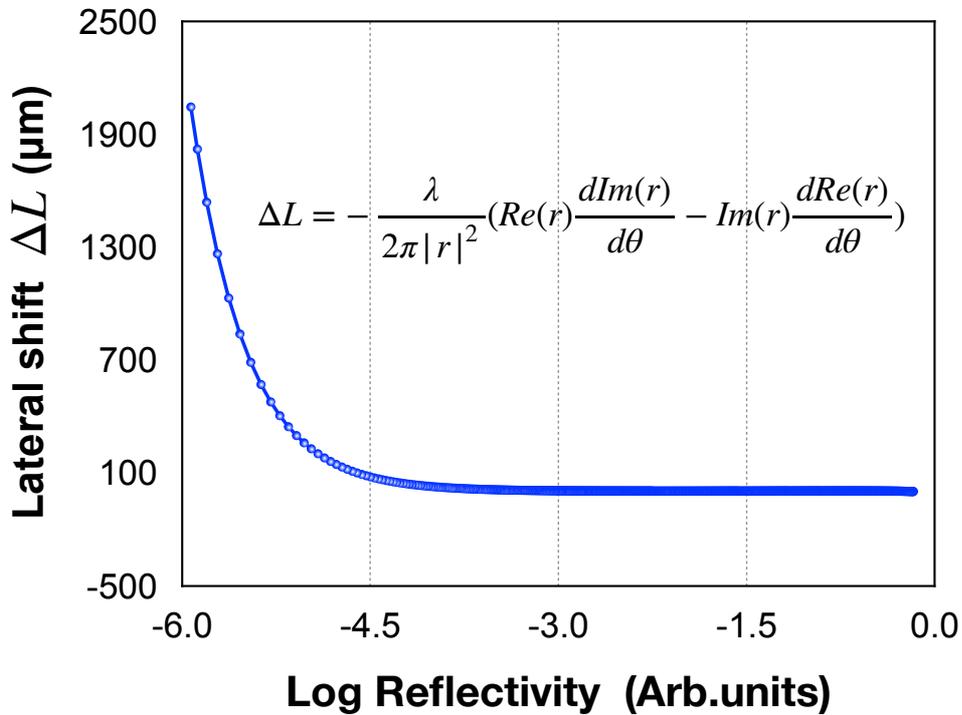

$$\Delta L = -\frac{\lambda}{2\pi |r|^2}(Re(r)\frac{dIm(r)}{d\theta} - Im(r)\frac{dRe(r)}{d\theta})$$

**Figure S1.** Relationship between plasmonic-based lateral position shift and reflectivity.

## 1.2. Finite element analysis (FEA)

We use finite element analysis (FEA) (COMSOL Multiphysics 5.2) to study the electric field distribution on this 2D GST-on-Au sensing substrate at the resonance angle.

As shown in **Figure S2** (a), a large electric field enhancement occurs at the sensing interface when surface plasmon resonance is excited. The resonance also results in a minimum reflectance and an enhanced lateral position shift in the reflected beam. We have conducted a comparison between the reflectivity and the lateral position shift of the Au-only substrate and our 2D GST-on-Au nanomaterial based substrate in **Figure S2** (b)(c). In both cases, the largest lateral position shift coincides with the minimum reflectivity point. An importance outcome of our analysis is that the addition of atomically thin GST material leads to a much deeper resonance dip. The corresponding maximum lateral position shift is 2107.33 μm, which is nearly 100 times larger than that associated with the case of using Au-only substrate. We can therefore assert that the atomically thin 2D GST layer will offer a superior sensitivity enhancement.

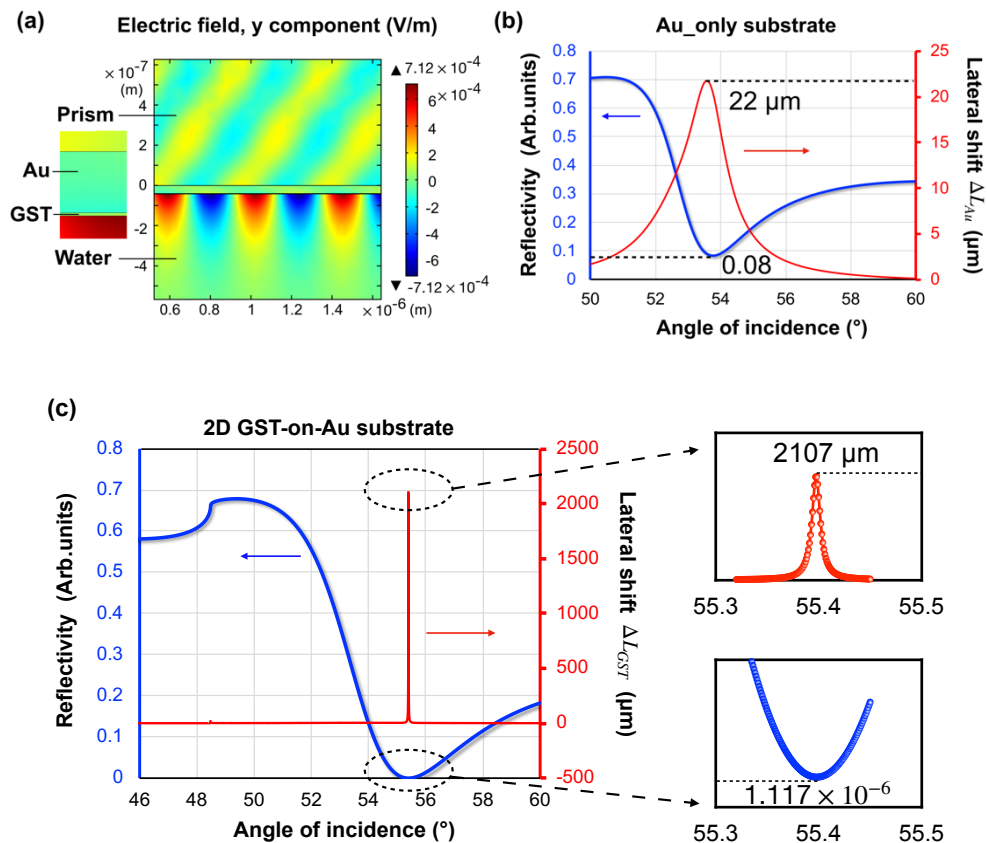

**Figure S2.** (a) Electric field distribution at SPR resonance angle 55.398°. (b) Simulation results of reflectivity and lateral position shift of Au-only substrate. (c) Simulation results of reflectivity and lateral position shift of 2D GST-on-Au substrate with zoom-in figures on the right showing the largest lateral position shift and lowest reflectivity.

## 2. Experimental

### 2.1. Optical characterization

The dielectric constant of GST in relation to the photon energy was measured through spectroscopic ellipsometry. For our experimental configuration, which uses a He-Ne laser (632.8 nm), the dielectric constant was determined to be 13.00+11.10i.

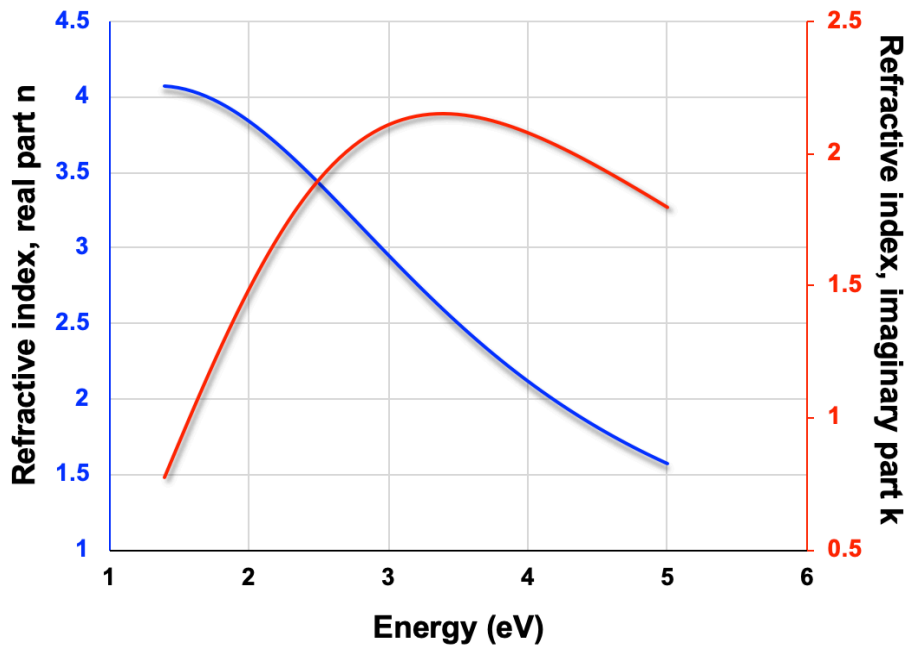

**Figure S3.** Refractive index measurement (real part and imaginary part) of GST under different photon energies.

### 2.2. 2D GST enhanced plasmonic sensing measurements

#### 2.2.1. Angular scanning reflectivity spectra in air

To evaluate the performance of the atomically thin GST-on-Au sensing substrate, we first measured its angular scanning reflectivity spectra in air. As a comparison, we also measured the Au-only substrate. The experimental results show good agreement with theoretical calculations, which confirms the reliability of our device and serves as a good calibration for the assessment of sensing performance. As shown in the **Figure S4**, the presence of GST material clearly leads to a deeper resonance dip (minimum intensity lowered by 50%).

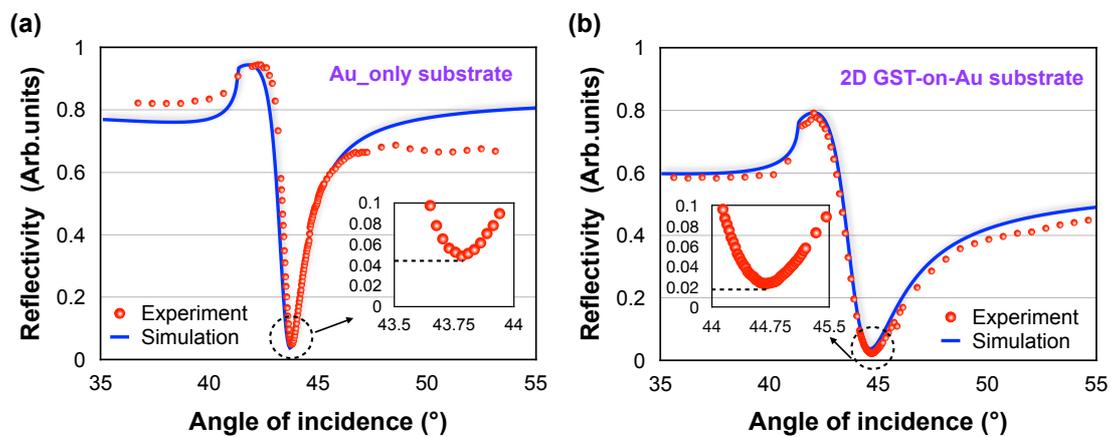

**Figure S4. Reflectivity spectra measurement.** (a) Au-only substrate (b) 2D GST-on-Au substrate.

### 2.2.2. Standard sensor evaluation using glycerol solutions

In our experimental setup, the prism (SF11 glass, Edmund Optics) with a refractive index of 1.7786 was mounted on a high-precision rotation stage (PR01/M, Thorlabs) with the beam fixed at the surface plasmon resonance dip angle for achieving maximum lateral position shift. The microfluidic chamber containing sample solution was designed with the size of 10mm*10mm*0.5mm. The orientation of the detection screen (2D Lateral Effect Position Sensor, PDP90A, equipped with a K-Cube PSD Auto Aligner, KPA101, Thorlabs) is in a head-on direction towards the light path. The incident light beam from a He-Ne laser (632.8 nm, Newport) is split into *p*-polarized and *s*-polarized light beams through a polarized beam splitter (Thorlabs). As a standard sensor evaluation procedure, glycerol solutions of different concentration levels were injected into the microfluidic chamber. **Figure S5** shows the signals for both *s*-polarized light and *p*-polarized light acquired under different glycerol concentrations using atomically thin GST-on-Au substrate. As shown in the plot, the position of p-polarized reflected light changes drastically while s-polarized reflected light remains in the same position upon injecting the glycerol/water solutions into the chamber. The signal is very stable due to the use of differential measurement scheme.

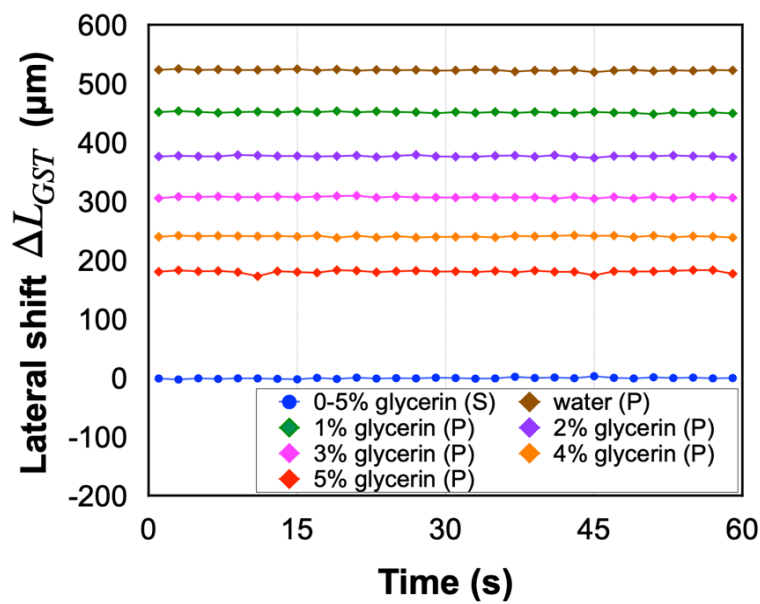

**Figure S5.** Lateral position shift of *p*-polarized and *s*-polarized light beam for different concentration levels of glycerol.

### 2.2.3. Biomolecule (BSA) sensing performance

The real-time biosensing capability of our device is also demonstrated through monitoring binding of BSA biomolecules, which have relatively high molecular weight (66463 Da), at different concentration levels. Solutions of the biomolecules with different concentrations ranging from 10 fM to 10 μM were detected and recorded in **Figure S6** based on lateral position shift measurement, which shows a linear increase in lateral position shifts with increasing BSA concentrations. Signal saturation will start when the concentration goes above $10^{-6}$ mol/L.

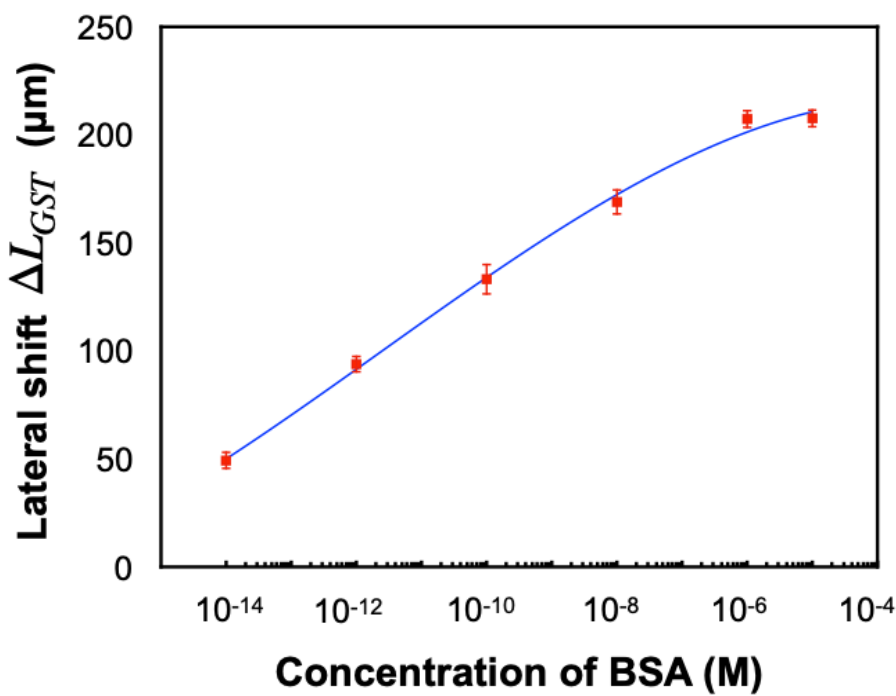

**Figure S6.** Detection of BSA molecules based on lateral position shift.

### 2.2.4. Detection of non-specific binding

To demonstrate the specificity of our sensing device, the non-specific binding between TNF-alpha and BSA molecules has also been detected in comparison with the specific antibody-antigen binding. We carried out TNF-α (tumor necrosis factor α) antigen detection using a sandwich immunoassay strategy. After flowing antigen containing solutions to the sensing substrate coated with antibody solutions, we further injected the capture antibody - monoclonal anti-TNF antibody to the sensing substrate. The lateral position shift signal can be increased to 16.70 μm when flowing 10 pM antibodies. As a negative control experiment, we use BSA as the control antibody [1-4]. As shown in **Figure S7**, the lateral position shift signal change when flowing BSA with a large concentration ($10^5$ times higher than anti-TNF antibody) is much smaller compared to flowing antibody, which shows the high specificity of our sensing device.

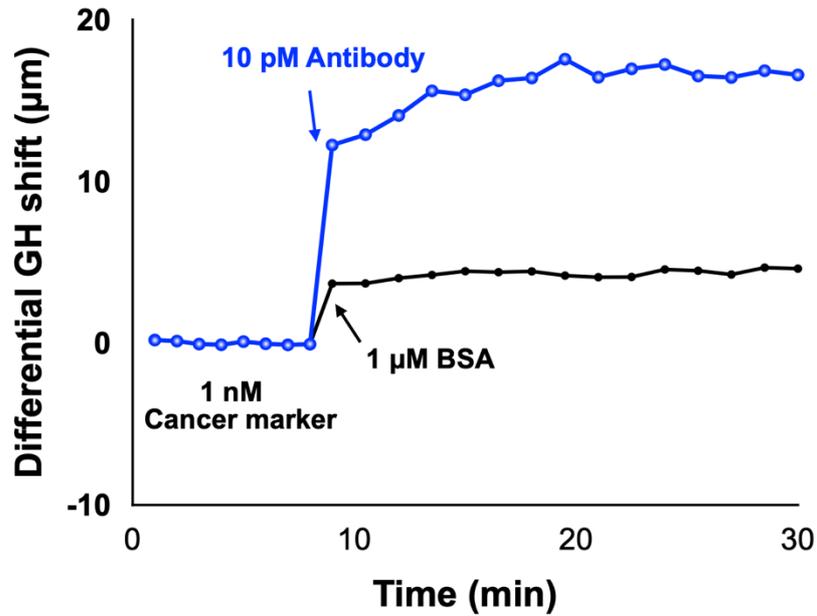

**Figure S7.** Detection of specific and non-specific binding based on lateral position shift. Blue curve shows the signal change when flowing capture antibody - monoclonal anti-TNF antibody to the sensing substrate while the black curve shows the signal change of flowing BSA as control antibody.

### 2.2.5. Cancer marker detection

The differential lateral position shift signals acquired during TNF-α detection with replicate measurements were summarized. The lateral position shift signal can reach 6.52 μm for 1 fm cancer marker and 53.15 μm for 1 nm cancer marker using GST-on-Au substrate.

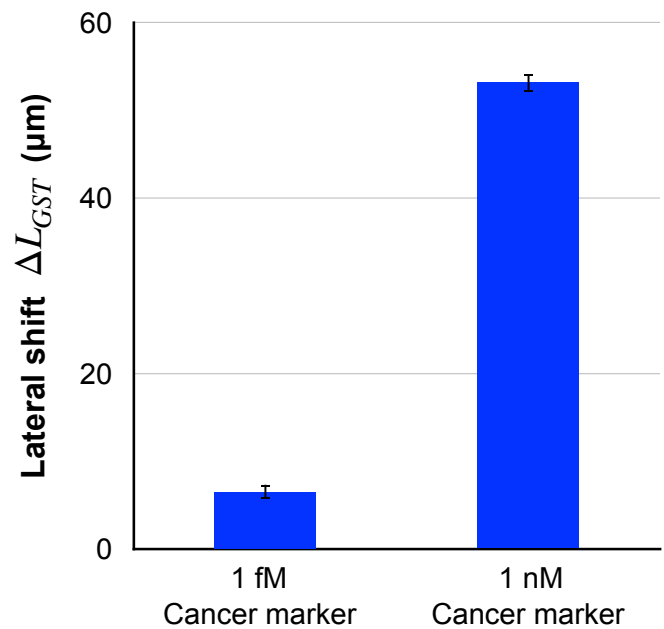

**Figure S8**. Detection of TNF-α based on lateral position shift.